\documentstyle[12pt,a4]{article}

\newcommand{\beq}[1]{\begin{equation}\label{#1}}
\newcommand{\eeq}{\end{equation}}

\begin{document}
\begin{titlepage}
\hfill{Freiburg THEP-96/1}

\hfill{gr-qc/9602033}
\vspace{1.5cm}

\begin{center}
{\huge Ehrenfest's Principle and the Problem of Time in Quantum Gravity}
\bigskip
\bigskip
\bigskip
\bigskip

T.\ Brotz\footnote[2]{Electronic address: brotz@phyq1.physik.uni-freiburg.de}
and C.\ Kiefer\footnote[3]{Electronic address: 
kiefer@phyq1.physik.uni-freiburg.de}
\bigskip

{\it Fakult\"at f\"ur Physik, Universit\"at Freiburg, Hermann-Herder-Str.3,}
\newline
{\it D-79104 Freiburg i.Br., Germany}
\bigskip
\bigskip
\bigskip
\bigskip


{\bf Abstract}
\bigskip

\begin{quote}
We elaborate on a proposal made by Greensite and others to solve the problem
of time in quantum gravity. The proposal states that a viable concept of
time and a sensible inner product can be found from the demand for the 
Ehrenfest
equations to hold in quantum gravity. We derive and discuss in detail exact
consistency conditions from both Ehrenfest equations as well as from the
semiclassical approximation. We also discuss consistency conditions arising
from the full field theory. We find that only a very restricted class of
solutions to the Wheeler-DeWitt equation fulfills all consistency conditions.
We conclude that therefore this proposal must either be abandoned as
 a means to
solve the problem of time or, alternatively, be used as an additional
boundary condition to select physical solutions from the Wheeler-DeWitt
equation.
\end{quote}
\bigskip
\bigskip

{\it Submitted to Nuclear Physics B}
\end{center}
\end{titlepage}


\section{Introduction}
\bigskip

Despite many attempts, a viable quantum theory of gravity is still elusive.
Apart from the absence of experimental hints, a major difficulty is the lack
of any obvious physical principle, which would enable one
to find such a theory, in analogy to
the role that the equivalence principle played in the construction of
general relativity.
\medskip

One candidate for such a guiding principle may be the ``problem of time in
quantum gravity'' which gained considerable interest recently (see, for
example, \cite{1}, \cite{2}, and Chap.6 of \cite{3}).
 This problem occurs in all systems whose
classical version is invariant under reparametrisations of the time parameter,
which leads to the absence of this parameter at the quantum level. The formal
question is how to handle the classical Hamiltonian constraint, $H \! \approx
\! 0$,
in the quantum theory. Connected with the problem of time is the ``Hilbert
space problem'' \cite{1,2} -- it is not at all obvious which inner product 
of states
one has to use in quantum gravity, and whether there is a need for such a
structure at all. It is thus also not clear whether there is any sensible 
notion
of unitary evolution of quantum states.
\medskip

Basically, the approaches to address the problem of time can be classified
as to whether an appropriate time variable is identifiable already at the
classical level, or only after quantisation. The former try, for example,
to cast the Hamiltonian constraint through an appropriate canonical
transformation into the reduced form $P_{_T} + h \approx 0$, where $P_{_T}$
denotes the momentum conjugate to the time variable $T$. If this were
possible, the problems of time and Hilbert space would be solved, since
the new form of the constraints would be transformed into a (functional)
Schr\"odinger equation upon quantisation, and the standard inner product
could be used. This has been shown to work in special examples \cite{2,4}, 
but it is
far from clear to which extent it works in the general case. It is, however,
known that it cannot work for the full configuration space \cite{4b}.
\medskip

Attempts to isolate a concept of time at the quantum level can be subdivided
into many possibilities \cite{1,2}, but have the common feature that time  --
if it exists at all at the most fundamental level -- has to be searched for
amongst the {\it dynamical} variables of the theory. Most approaches implement
the classical constraint \`{a} la Dirac as a condition on physically allowed
wave functionals, $\hat{H} | \Psi \rangle_{phys}= 0 $, the Wheeler-DeWitt
equation. One may then, motivated by the indefinite kinetic term in $\hat{H}$,
use a Klein-Gordon type of inner product \cite{5}.
Since this inner product is not positive definite, however, many problems 
arise which have
provoked some authors to invoke a ``third quantisation'' of the theory (see
the review in \cite{1,2}).
A more recent attempt consists in constructing a positive definite physical
Hilbert space from some auxiliary Hilbert space via some ``spectral analysis
proposal'' \cite{6}.
\medskip

A necessary requirement for all approaches is of course the ability to
recover a sensible notion of semiclassical time from the full theory.
This is mostly done in the context of some Born-Oppenheimer type of
approxima\-tion scheme which employs an expansion of the full wave functional
in powers of the Planck mass (see, e.g., the review in \cite{7}).
A notion of semiclassical time, or ``WKB time'', emerges thereby from the
{\it phase} of the wave functional in the first order of
approximation. Higher orders modify the definition of WKB time through
the {\it back reaction} of the quantum degrees of freedom onto the
semiclassical ones.

A semiclassical approximation of a somewhat different kind can also be made
for the approach in \cite{6}, where an explicit map between the physical 
Hilbert
space of the full theory and the standard ``external Hilbert space'' of the
Schr\"odinger equation can be made through the use of ``almost ideal clocks''
\cite{8}.
\medskip

An interesting mixture of some of the above ideas has been suggested by
Greensite \cite{9,10} (see also similar attempts in \cite{11,12}). Contrary to
the semiclassical approximation, time is there defined by the {\it exact phase}
of the {\it full} wave functional, and it is insisted on the validity of 
an exact
Schr\"odinger-type of inner product. Since it would be inconsistent to demand
the validity of a Schr\"odinger equation for the full theory, the weaker 
condition of Ehrenfest
equations for appropriate expectation values is imposed. The idea then is to
{\it define} time by the validity of these equations. We call this notion
``Ehrenfest time'' and reserve the label ``phase time'' for the more general
concept of defining time from the full wave functional without further
conditions. The idea implicit in [10-13] is that a notion of ``phase time''
automatically implies an ``Ehrenfest time''.
\medskip

The purpose of our work is to investigate in detail how far the notion of
Ehrenfest time can be used as a viable candidate for a concept of time in
quantum gravity. We shall find that, contrary to what has been thought
previously, only a small subset of solutions to
$H\Psi = 0$ allows the definition of an Ehrenfest time. Moreover, only
a small subset of these have a sensible semiclassical limit. Our overall
conclusion will thus be that one has either to reject this proposal as
a way to solve the problem of time in quantum gravity or, alternatively, to
use it as an additional boundary condition to {\it select} sensible
wavefunctions from all solutions to the Wheeler-DeWitt equation.
\bigskip

Our paper is organised as follows. In Section 2 we start with a review
of the proposal put forward in \cite{9,10}. We then explore further properties
of the first Ehrenfest equation (the one associated with the configuration
variables) and demonstrate, in particular, that a phase time does not
necessarily possess the properties of an Ehrenfest time. We then derive
an exact condition for the validity of the second Ehrenfest equation
(the one associated with the canonical momentum).

Section 3 ist devoted to a detailed comparison with the semiclassical
approximation. We shall show, first, that the solutions which lead to an
Ehrenfest time do not, in general, allow a sensible semiclassical limit.
We then study the connection with the back reaction-corrected WKB time in
those cases where a sensible semiclassical limit does hold.

Section 4 investigates particular issues which appear in the full,
infinite-dimensional, theory. {\it If} a solution to the Wheeler-DeWitt
equation admits a (local) Ehrenfest time, we show that it can be mapped
to a scalar function on superspace (the configuration space of three-metrics
modulo diffeomorphisms). We can also conclude that this result must hold for
all orders of the semiclassical approximation scheme discribed in \cite{7}.

Section 5 presents our conclusion, in particular a discussion of
the above proposal in the light of the ``problems of time'' in \cite{2}.

Some lengthy calculations are relegated to an appendix.


\section{Ehrenfest's equations in quantum gravity}
\bigskip

In this section we investigate the consequences which arise from the demand
for the validity of Ehrenfest's equations in quantum gravity. We start with
a review of the proposal in \cite{9,10} and then proceed to work out further
properties.
\medskip

The coordinates in configuration space shall be denoted by $\{ q^\alpha \} $.
In quantum
general relativity, these are the coordinates on full superspace. However,
in practice it is more convenient to work with the (redundant) variables
before the diffeomorphism group has been factored out -- the three-metric
and matter fields. Moreover, we restrict ourselves here to finite-dimensional
models (thus, $\alpha $ runs from $1,...,N$) and discuss some peculiarities of
infinite dimension in Section 4.
\medskip

The proposal is to find a coordinate transformation $\{q^\alpha\} \! 
\rightarrow
\! \{t, \eta^i\}$, such that for an ``observable'' $A(q^\alpha,p_\beta)$  
Ehrenfest's condition holds:
\beq{1}
\frac{d}{dt} <A> = \int D\eta \, M(t,\eta) \; \Psi^*(t,\eta) \, i[H,A] \,
                   \Psi(t,\eta) + <\frac{\partial A}{\partial t} > \; ,
\eeq
where
\beq{2}
<A> \equiv \int D\eta \, M(t,\eta) \Psi^*(t,\eta) \; A \; \Psi(t,\eta)
\eeq
is the standard expression for an expectation value in the Schr\"odinger-type
inner product. It is assumed therein that the full Hamiltonoperator, $H$, 
of the 
Wheeler-DeWitt equation $H\Psi = 0$ can be used as the ``dynamical operator''
describing the evolution of expectation values. The important point is that a 
time variable is singled out by
this prescription, since the integration is over the $\eta$-variables only.
The wave function $\Psi$ in these expressions is assumed to be a {\it given}
solution to the Wheeler-DeWitt equation; $M(t,\eta)$ denotes a measure which 
is specified below.
\medskip

Note that the ``observables'' $A$ are {\it not} assumed to commute with the 
full Hamiltonian, since otherwise the content of (\ref{1}) would be trivial.
They are ``observables'' in the sense of \cite{13} 
(in contrast to ``perennials'') -- quantities which commute with the 
diffeomorphism constraints
of general relativity, but not necessarily with the Hamiltonian constraint.
\medskip

The Hamiltonian $H$ can formally be written as (we use $\eta$ with greek
indices if $t$ is adjoined to ${\eta^i}, \eta^0 \equiv t$)
\beq{3}
H =\, ''\, G'^{\alpha \beta} \frac{\partial }{\partial \eta^\alpha}
      \frac{\partial }{\partial \eta^\beta} \; '' + \xi {\cal R} + V(t,\eta)
\; ,
\eeq
where the quotations in the kinetic term emphasise that the factor-ordering
problem has not been addressed yet, and where an additional factor-ordering
ambiguity proportional to the Ricci scalar of configuration space has been
taken into account. The supermetric with respect to the $(t,\eta)$-coordinates
is labeled with a prime.
\medskip

The task is now to explicitly determine the above coordinate transformation
on configuration space. Consider first the Ehrenfest equation with respect
to the configuration variables, i.e., choose $A=\eta^i$. We
shall refer to this as the ``first Ehrenfest equation''. One finds
\beq{4}
[H,\eta^i] \Psi = -2i G'^{\alpha \beta} \delta_{\alpha}^{\; i}
                  \frac{\partial }{\partial \eta^\beta} \Psi \; .
\eeq
The demand for $H$ to be hermitean requires that $G'^{0i}$ be zero;
otherwise one would be left with $t$-derivatives which cannot be
transferred from $\Psi$ to $\Psi^*$ upon partial integration in the above
inner product. The condition $G'^{0i}=0$ means that the time direction runs
orthogonally to hypersurfaces of constant time.
If we demand, in addition, the invariance of the first Ehrenfest equation
with respect to transformations of the variables $\{ \eta^i \}$  
on $t=const.$, the measure in (\ref{2}) is fixed to be $M(t,\eta) = 
\sqrt{G'} $, where $G'$ denotes
the determinant of the metric $G'_{\alpha \beta}$, and the kinetic term in 
(\ref{3}) is the Laplace-Beltrami-operator.
The Hamiltonian (\ref{3}) thus can be written as

\begin{eqnarray} \label{5}
 H & = &-\frac{1}{2} \left[\frac{1}{\sqrt{G'}} \frac{\partial}{\partial \eta^i}
\sqrt{G'} G'^{ij} \frac{\partial}{\partial \eta^j} + \frac{1}{\sqrt{G'}}
\frac{\partial}{\partial t} \sqrt{G'} G'^{00} \frac{\partial}{\partial t}
 \right] + \xi {\cal R} + V(t,\eta) \nonumber \\
  & =: &  -\frac{1}{2} D^2 -\frac{1}{2} D_0^2 +  \xi {\cal R} + V(t,\eta) \; .
\end{eqnarray}
It was shown in \cite{10} that for the first Ehrenfest equation to be 
fulfilled, $\Psi $ must be of the form
\beq{6}
i\frac{\partial \Psi}{\partial t} = \left( -\frac{1}{2} D^2 + \widetilde{V} -
\frac{i}{2} \frac{\partial \ln \sqrt{G'}}{\partial t} \right) \Psi \; .
\eeq
Despite its formal similarity, this is {\it not} a Schr\"odinger equation,
since $\Psi$ is here assumed to be a given solution to the Wheeler-DeWitt
equation. Decomposing $\Psi$ into its real and imaginary parts and using
$H \Psi=0$, one finds from (\ref{6}) that
\beq{7}
\widetilde{V}= \frac{1}{\Psi^* \Psi} \left[ \frac{1}{2} Re(\Psi^* D^2 \Psi) -
Im(\Psi^* \partial_t \Psi) \right] \;,
\eeq
\beq{8}
\partial_t \ln \sqrt{G'} =  \frac{1}{\Psi^* \Psi} \left[Im(\Psi^* D_o^2 \Psi )
- \partial_t (\Psi^* \Psi) \right] \;.
\eeq
Writing $\Psi$ as
\beq{9}
 \Psi(t,\eta) =  \varrho(t,\eta) e^{i\theta(t,\eta)}\; ,
\eeq
one finds from (\ref{8})
\beq{10}
G'^{00} \partial_t \theta = 1+ \frac{f(\eta)}{\sqrt{G'}\varrho^2}
                          \equiv \kappa (t,\eta)
\eeq
as the condition which must be satisfied after employing the coordinate 
transformation
$\{q^{\alpha} \} \rightarrow \{t,\eta^i \}$. Note that $f=0$ was chosen in
\cite{10}. Note also that the validity of the first Ehrenfest equation includes
conservation of probability, since $\frac{d}{dt}<\!\Psi|\Psi\!>=0$  follows for
$A \equiv 1\!\!1$. As can be immediately recognised from (\ref{10}),
the proposal works only for {\it complex} wave functions, i.e. for
$\theta \neq 0$.
\medskip

It is instructive to make a connection between Ehrenfest time and derivatives
with respect to the
old coordinates $\{q^\alpha\}$. Writing
\[ \frac{\partial }{\partial t} = T^\alpha \frac{\partial }{\partial q^\alpha},
\]
we find with the help of (\ref{10})
\[\frac{\partial \theta}{\partial t}
   = T^{\alpha} \frac{\partial \theta}{\partial q^\alpha}
   = \kappa \; G'_{00}
   = \kappa \; G_{\alpha \beta} \frac{\partial q^\alpha}{\partial t} 
     \frac{\partial q^\beta}{\partial t}
   = \kappa \; G_{\alpha \beta} T^{\alpha} T^\beta \; ,\]
where $G_{\alpha \beta}$ denotes the components of the configuration space
metric with respect to $\{q^\alpha\}$. Since this yields $T^\beta = \kappa^{-1}
G^{\alpha \beta} \, \partial\theta / \partial q^\alpha$, one has
\beq{11}
\frac{\partial }{\partial t} = \kappa^{-1} G^{\alpha \beta} 
\frac{\partial \theta}{\partial q^\alpha}
       \frac{\partial }{\partial q^\beta}\; .
\eeq
This expression shows explicitly that Ehrenfest time is constructed from the
phase of $\Psi$, i.e., that it is proportional to phase time. This also
resembles the definition of the WKB time in the semiclassical approximation
(see below). {}From $G'_{0i}=0$ and (\ref{11}) one finds that 
$\partial\theta/\partial \eta^i=0$ and, thus, $\theta$ is a functional
of $t$ only. 

Decomposing now the Wheeler-DeWitt equation into its real and imaginary parts,
one then finds (a dot denoting a derivative with respect to $t$): 
\medskip

\noindent
Real part:
\beq{12}
\frac{1}{2} G'^{00} \dot{\theta}^2 + \frac{1}{\varrho}H\varrho = 0 \; ,
\eeq
Imaginary part:
\[
\frac{1}{2} \frac{1}{\sqrt{G'}}\left(\partial_t \sqrt{G'} G'^{00}\right)
\dot{\theta}+ G'^{00} \frac{\dot{\varrho}}{\varrho}\dot{\theta} 
+ \frac{1}{2} G'^{00} \ddot{\theta}=0 \]
\[ \Leftrightarrow \;\;\;\;
\frac{1}{2} \partial_t \ln \sqrt{G'}\left[1 + \frac{f(\eta)}{\sqrt{G'}
\varrho^2}\right] + \frac{1}{2} \partial_t \left[1 + 
\frac{f(\eta)}{\sqrt{G'}\varrho^2}\right]
+ \frac{\dot{\varrho}}{\varrho}\left[ 1+ \frac{f(\eta)}{\sqrt{G'}\varrho^2}
\right] =0
\]
\beq{13}
 \Leftrightarrow  \hspace{1cm} \partial_t (\sqrt{G'} \varrho^2) = 0 \;.
\eeq
Equation (\ref{13}) means that the solution $\Psi$ is {\it stationary} 
with respect to 
Ehrenfest time, i.e., the integrand in $<\Psi|\Psi>$, which is given by
$\sqrt{G'} \Psi^*\Psi $, is itself time independent. Therefore,
$G'^{00} \dot{\theta} = \kappa(\eta)$ from (10). Consequently,
\[ \frac{d}{dt} < \eta^i >  = \frac{d}{dt} \int D\eta \sqrt{G'} \varrho^2 
\eta^i = 0 \; , \]
and thus
\beq{14}
 \frac{d}{dt} <A(t,\eta)> = < \frac{\partial}{\partial t} A(t,\eta)>,
\eeq
i.e., the time dependence in $<\!A(q^\alpha)\!>$ arises solely from the 
explicit
dependence of $A$ on time. Since therefore 
$0=\frac{d}{dt} <\eta^i> = <i[H,\eta^i]>$,
one could call the variables $\eta^i$ {\it perennials} \cite{13} with respect
to the scalar product used here.

We note that in the semiclassical approximation (see Sect.3) equation
(\ref{13})  just corresponds to the ``prefactor equation'' in the highest
order of approximation when written in ``comoving coordinates'' \cite{2,7,14}.
The above derivation shows that $\Psi$ has the form
$ \Psi = G'(t,\eta)^{-1/4} g(\eta) e^{i \theta(t)} $ with some function $g$ 
depending on $\eta$ only.
\medskip

We now address the conditions which are necessary for the second Ehrenfest
equation (see equation (\ref{16}) below) to be fulfilled. In \cite{10} an 
approximate validity was shown in the
case where the phase is rapidly varying. Here we shall present an exact 
condition.

As in quantum field theory on curved backgrounds we shall use the momentum
$p_i$ which is hermitean with respect to the measure $\sqrt{G'}$ in the inner
product. It reads
\beq{15}
p_i  = -i  G'^{-1/4} \partial_i G'^{1/4}
     =  -i \partial_i - \frac{i}{2} \left( \partial_i \ln\sqrt{G'} \right) \;.
\eeq
We demand that
\beq{16}
\frac{d}{dt} <p_i> = i <[H, p_i]> + <\frac{\partial p_i}{\partial t}> \; .
\eeq
The calculations are straightforward, but somewhat lengthy,  and have
thus been relegated to the appendix.
The result is that (\ref{16}) can only hold if the last term vanishes
explicitly, i.e., if
\beq{17}
< \frac{\partial p_i}{\partial t} > = 0 = <\partial_i\partial_t \ln \sqrt{G'}
 > \;.
\eeq
This yields an additional restriction on allowed physical states. Note that
-- in contrast to $<\eta^i>$ -- stationarity of $\Psi$ does {\it not}
lead to $\frac{d}{dt}<p_i> =0$.
\bigskip

We conclude this section with a simple example which demonstrates that not
every ``phase time'' satisfies the condition for an Ehrenfest time.

Consider the  model of an indefinite harmonic oscillator (arising, e.g.,
from a Friedmann model with a conformally coupled scalar field \cite{15}).
The Wheeler-DeWitt equation reads
\[ \left[ \frac{\partial^2 }{\partial x^2}
  - \frac{\partial^2}{\partial y^2} -x^2 +y^2 \right]\Psi(x,y) = 0 ,\]
and we shall choose in particular the simple solution
\[ \Psi= \exp(ixy) \;.\]
{}From (\ref{11}) with $\kappa=1$ (this choice is justified in Sect.3)
we have
\[ \frac{\partial}{\partial t} = - y \frac{\partial }{\partial x}
    + x \frac{\partial }{\partial y}\;, \]
yielding $\dot{x}= -y$, $\dot{y}=x$, and $\dot{\theta}=x^2-y^2$. Since
$\theta $ must not depend on $\eta$ (see above), we find from the condition
$\partial_\eta \theta=0$, $\partial_\eta \dot{\theta} =0$ the equations
(a prime denotes a derivative with respect to $\eta$ )
\[ y x'+xy' = 0 \; ,\hspace{1cm}  xx' - yy' = 0 \;, \]
which allow only the trivial solution $x'\!=\!y'\!=\!0$. Thus, there does
not exist
any coordinate transformation from $(x,y)$ to $(t,\eta)$, such that $t$ has
the properties of an Ehrenfest time (independent of the validity of the
second Ehrenfest equation).
The demand for the Ehrenfest equation to hold is thus much more restrictive
than was originally assumed [10-13].

On the other hand, if one reversed, e.g., the sign of the potential in this
example and considered a solution of the type $\Psi = \exp \left(\frac{i}{2}
(x^2-y^2)\right)$, the Ehrenfest conditions could be fulfilled.


\section{Ehrenfest time and the semiclassical approximation }
\bigskip

In this section we make a detailed comparison of the Ehrenfest time with
the standard semiclassical approximation to quantum gravity and the
approximate notion of WKB time \cite{7}.
\medskip

First, we note from (\ref{11}) that in order to be able to recover the
momentum in the semiclassical limit through $p_\alpha = \partial \theta /
\partial q^\alpha $, we must choose $\kappa = const$ (for simplicity
we choose $\kappa=1$). Next, using (\ref{11}) with $\kappa=1$, we write
the ``Ehrenfest'' condition (\ref{10}) in the form
\beq{18}
\frac{1}{2} G^{\alpha \beta} \frac{\partial \theta}{\partial q^\alpha}
\frac{\partial \theta}{\partial q^\beta}  - \frac{1}{2G'^{00}} = 0 \; .
\eeq
If a semiclassical approximation were valid, $\theta$ would also obey the
Hamilton-Jacobi equation
\beq{19}
\frac{1}{2} G^{\alpha \beta} \frac{\partial \theta}{\partial q^\alpha}
\frac{\partial \theta}{\partial q^\beta}  + V \approx 0 \; .
\eeq
Since $G'^{00}$ can be a function of $t$ only (see (\ref{10}) with
$\theta=\theta(t)$ and $\kappa=1$), the last two equations would only 
be compatible if
$V\approx V(t)$. Since $V$ is a given function, this is of course a strong 
restriction on the class of allowed semiclassical states.
\medskip

To discuss the connection with the semiclassical approximation reviewed in
\cite{7}, we start from the Wheeler-DeWitt equation
\beq{20}
 \left( - \frac{1}{2M} \frac{1}{\sqrt{G}} \frac{\partial}{\partial h^a }
 \sqrt{G} G^{ab} \frac{\partial}{\partial h^b}
 + MV + {\cal H}_m \right) \Psi = 0
\eeq
with
\beq{21}
 {\cal H}_m = \frac{1}{2} \left( -\frac{1}{\sqrt{h}} \frac{\partial^2}{
 \partial \phi^2} + \sqrt{h}(m^2\phi^2+  U(\phi) ) \right)
\eeq
being the Hamiltonian for a homogeneous scalar field. The variables $\{h^a \}$
denote the components of the three-dimensional metric, and $h$ is its 
determinant.

The important point to note is that the degrees of freedom $\{q^\alpha\}$
have been divided into some ``heavy ones'' with large ``mass'' $M$
(the gravitational degrees of freedom)
and some ``light'' degree of freedom, the scalar field $\phi$ (see \cite{7}).
Equation (\ref{18}) then reads
\beq{22}
\frac{1}{2M} G^{ab} \frac{\partial \theta}{\partial h^a }
     \frac{\partial \theta}{\partial h^b} + \frac{1}{2\sqrt{h}}
     \left( \frac{\partial \theta}{\partial \phi} \right)^2
     - \frac{1}{2G'^{00}} = 0 \; .
\eeq
Expanding now the phase $\theta$ into inverse powers of $M$,
\beq{23}
\theta = M S_0 + Re(S_1) + M^{-1}Re(S_2)+ ... \; ,
\eeq
Equation (\ref{22}) yields the following equations at consecutive orders of
$M$:
\smallskip

\noindent
$O(M^2)$:
\beq{24}
  \frac{1}{\sqrt{h}} \left( \frac{\partial S_0}{\partial \phi}\right)^2
  -\left. \frac{1}{G'^{00}}\right|_{M^2} = 0 \;,
\eeq
$O(M^1)$:
\beq{25}
  G^{ab} \frac{\partial S_0}{\partial h^a} \frac{\partial S_0}{\partial h^b }
  + \frac{2}{\sqrt{h}} \frac{\partial S_0}{\partial \phi}
        \frac{\partial Re(S_1)}{\partial \phi}
  - \left. \frac{1}{G'^{00}} \right|_{M^1} = 0 \;,
\eeq
$O(M^0)$:
\beq{26}
    2G^{ab} \frac{\partial S_0}{\partial h^a} \frac{\partial Re(S_1)
       }{\partial h^b }
    + \frac{1}{\sqrt{h}} \left(\frac{\partial Re(S_1)}{\partial
       \phi}\right)^2
    + \frac{2}{\sqrt{h}} \frac{\partial S_0}{\partial \phi}
      \frac{\partial Re(S_2)}{\partial \phi}
    - \left. \frac{1}{G'^{00}} \right|_{M^0} = 0 \;.
\eeq
On the other hand, the standard Born-Oppenheimer type of expansion yields 
\cite{7}
\smallskip

\noindent
$O(M^2)$:
\beq{27}
 \frac{\partial S_0}{\partial \phi} = 0 \; ,
\eeq
$O(M^1)$:
\beq{28}
 \frac{1}{2} G^{ab} \frac{\partial S_0}{\partial h^a} \frac{\partial S_0}{
\partial h^b} + V = 0 \; ,
\eeq
$O(M^0)$:
\beq{29}
 G^{ab} \frac{\partial S_0}{\partial h^a} \frac{\partial D}{\partial h^b}
= \frac{1}{2} G^{ab} \frac{\partial^2 S_0}{\partial h^a \partial h^b} D,
\eeq
\beq{30}
 i G^{ab} \frac{\partial S_0}{\partial h^a} \frac{\partial \chi}{\partial h^b}
 \equiv i \frac{\partial \chi}{\partial t_{_{W\!K\!B}}}
 = {\cal H}_m \chi \;.
\eeq
Here we have introduced
\beq{31}
 \chi(h^a,\phi) \equiv D(h^a) \exp\{iS_1(h^a,\phi)\}
                \equiv \rho(h^a,\phi) \exp\{i Re(S_1) \}
\eeq
and chosen for $D(h^a)$ the usual prefactor equation \cite{7,14}. We can thus
write (\ref{30}) in the form
\beq{32}
 G^{ab} \frac{\partial S_0}{\partial h^a} \frac{\partial Re(S_1)}{\partial
 h^b} =
i G^{ab} \frac{1}{\rho}\frac{\partial S_0}{\partial h^a}
\frac{\partial \rho}{\partial h^b} - \frac{1}{\chi} {\cal H}_m \chi \; .
\eeq
Up to order $M^0$, the full wave function thus reads
\beq{33}
 \Psi = \frac{1}{D} e^{i M S_0 } \chi = \frac{\rho}{D}
        e^{i\left( M S_0 + Re(S_1)\right)}\;.
\eeq
Comparison of the two expansion schemes then yields
\begin{eqnarray*}
\left.\frac{1}{2G'^{00}}\right|_{M^2} & = &  0 \; , \\
\left.\frac{1}{2G'^{00}}\right|_{M^1} & = & -V \; , \\
\left.\frac{1}{2G'^{00}}\right|_{M^0} & = &
      i G^{ab}\frac{1}{\rho}\frac{\partial S_0}{\partial h^a}
      \frac{\partial \rho}{\partial h^b}
    - \frac{1}{\chi} {\cal H}_m \chi
    + \frac{1}{2} \frac{1}{\sqrt{h}} \left(\frac{\partial
       Re(S_1)}{\partial \phi} \right)^2  \;.
\end{eqnarray*}
Using these results in (\ref{22}), one gets
\beq{34}
 \frac{G^{ab}}{2M} \left( M \frac{\partial S_0}{\partial h^a}
 +\frac{\partial Re(S_1)}{\partial h^a} \right)
  \left( M \frac{\partial S_0}{\partial h^b}
 +\frac{\partial Re(S_1)}{\partial h^b} \right)
 + M V  + \frac{{\cal H}_m \chi}{\chi} 
\eeq
\[ - i  \frac{G^{ab}}{\rho} \frac{\partial S_0}{\partial h^a}
 \frac{\partial \rho}{\partial h^b} + O(\frac{1}{M}) = 0.
\]
Multiplying this equation with $\chi^*$ and integrating over $\phi$ yields
\beq{35}
\frac{G^{ab}}{2M} \left( M\frac{\partial S_0}{\partial h^a}
+ <\!\chi|\frac{\partial Re(S_1)}{\partial h^a} \chi\!>_{_\phi} \right)
   \left( M\frac{\partial S_0}{\partial h^b}
+ <\!\chi|\frac{\partial Re(S_1)}{\partial h^b} \chi\!>_{_\phi} \right)
+ M V 
\eeq
\[
+  <\!\chi|{\cal H}_m \chi\!>_{_\phi} + O(\frac{1}{M}) = 0\; ,
\]
where $<\!\psi|\varphi\!>_{_\phi} \equiv \int D\phi \; \psi^*\varphi$. In the
derivation of (\ref{35}) we have made use of the fact that $\chi$, which is
a solution to the Schr\"odinger eqution (\ref{30}), can be normalised, and
thus
\begin{eqnarray*}
<\! \chi| \frac{1}{\rho}G^{ab} \frac{\partial S_0}{\partial h^a}
   \frac{\partial \rho}{\partial h^b} \chi \!>_{_\phi} & = &
\int d\phi \; e^{-i Re(S_1)} G^{ab} \frac{\partial S_0}{\partial h^a}
   \frac{\partial \rho}{\partial h^b} \; \rho e^{i Re(S_1)} \\
& = & G^{ab} \frac{\partial S_0}{\partial h^a} \frac{\partial }{\partial h^b}
      \int d\phi \; \frac{1}{2} \rho^2\\
& = & 0 \; \; .
\end{eqnarray*}
Equation (\ref{35}) is just the ``back reaction corrected'' Hamilton-Jacobi
equation in this order of approximation \cite{7}, which we have here shown
to be consistent with a semiclassical approximation of the ``Ehrenfest
condition'' (\ref{22}). Note that, in contrast to the spirit of the standard
approach, the concept of time is here fixed once and for all by the Ehrenfest
condition, whereas in the standard approach a concept of time emerges at
$O(M^0)$ (the WKB time introduced in (\ref{30})) and is modified at higher
orders through back reaction effects \cite{7}. This modification -- which in
the standard approach is made by hand -- follows here automatically.
\medskip

The scalar product which we have been using in (\ref{35}) is not the one of
the full theory, see (\ref{2}). How are these inner products related?
\medskip

In the highest order of the semiclassical approximation, Ehrenfest time
agrees with WKB time, and the integration in the
total inner product is thus over $D\tilde{\eta} D\phi$, where $\tilde{\eta}$
denotes the part of the three-metric orthogonally to the flow generated by
WKB time. It was shown in \cite{14} that the choice of a prefactor $D$ 
(see (\ref{29})) which is {\it sharply peaked} in ${\tilde{\eta}}$ leads to
\[ \int D\tilde{\eta} \; D\phi \; \Psi^*\Psi \approx
   \left( \int D \tilde{\eta} \; D^{-2} \right) \left( \int D\phi \; 
\chi^* \chi
   \right) + O(M^{-1})\; .
\]
Since the left-hand side agrees with the full inner product
$\int D\eta \Psi^* \Psi $ up to order $O(M^{-1})$, the use of the
$\phi$-inner product in (\ref{35}) to find an effective equation for the
gravitational field is justified.
\medskip

We also note that one can somewhat modify the standard semiclassical
expansion to incorporate back reaction directly into the gravitational
part of the Wheeler-DeWitt equation \cite{16}.
\bigskip

Simple minisuperspace models may be used to explicitly compute the various
concepts of time \cite{17}.

In a ($k\!=\!-1$) Friedmann model with a scalar field, for example, the
Wheeler-DeWitt equation reads
\beq{36}
 \left[ \frac{1}{a}\frac{1}{M}\frac{\partial^2}{\partial a^2} +\frac{1}{a^2}
      \frac{1}{M}\frac{\partial}{\partial a}
 - \frac{1}{a^3}\frac{\partial^2 }{\partial \phi^2} +M a\right]
 \Psi(a,\phi) = 0 \;,
\eeq
where $a$ denotes the scale factor.
We choose the following solution at $O(M^0)$:
\[ \Psi(a,\phi)\approx a \exp{(-i\frac{Ma^2}{2})} \chi(a,\phi) \;, \]
where
\[ \chi(a,\phi) = \frac{1}{\pi^{1/4}} \frac{\sigma^{1/2}}{(\sigma^2
                      +i/2a^2)^{1/2}} \exp{(-\frac{\phi^2}{2\sigma^2+i/a^2})} 
\;,
\]
and $\sigma$ is an arbitrary constant. The WKB time is then given by
\[ \frac{\partial }{\partial t_{_{WKB}}} =
         G^{00} \frac{\partial S_0}{\partial a}\frac{\partial }{\partial a}
         =\frac{\partial }{\partial a},
\]
while for the ``back reaction corrected'' WKB time $\tilde{t} $ (which agrees
with the Ehrenfest time in this order of approximation after the $\phi$-field
has been integrated out) one has \cite{17}
\[
\frac{\partial}{\partial \tilde{t}} = \left( 1- \frac{1}{4M\sigma^2 a^4}
          \right) \frac{\partial }{\partial a} \;.
\]
Choosing an exact solution which reduces to the above semiclassical solution
for $a \rightarrow \infty$, one can find the following approximate 
``Ehrenfest time''
\[ \frac{\partial }{\partial t} = \left[ 1- \frac{2}{Ma^4} \left(
       \frac{1}{4\sigma^2} -\frac{\phi^2}{4 \sigma^4} - \frac{1}{M}
       \right) \right] \frac{\partial }{\partial a}
       + \frac{\phi}{2a^5\sigma^4} \frac{\partial}{\partial \phi}
\]
which of course, in contrast to $t_{_{WKB}}$ and $\tilde{t}$, is defined 
through {\it all} variables of the theory.

Another example of a WKB and a phase time -- there called 
Ehrenfest time in the spirit of \cite{10}, but in our sense  to be understood
as a candidate for an Ehrenfest time -- can be found in \cite{17b}.
\section{Consistency conditions from field theory}
\bigskip

Up to now, we have only considered models with a finite number of degrees of
freedom. Since the Ehrenfest proposal was intended to apply for full
quantum gravity, it must be investigated to which extent the above
conditions can be generalised to the field theoretic case. We shall proceed
analogously to \cite{18}, where consistency conditions were discussed for the
semiclassical approximation. While there it was found that the WKB time
cannot exist as a scalar function on the space of three-metrics, but only
on superspace, we shall here find an analogous result for the Ehrenfest
time. We shall thereby also be led to
a characteristic property of transformation to the Ehrenfest coordinates
$(t,\eta)$.
\medskip

The local form $\tau(x)$ of the phase time (\ref{11}) would read (recall that
$\kappa=1$)
\beq{37}
  \frac{\delta}{\delta \tau(x)} \equiv \xi_x = G^{\alpha \beta}  \frac{\delta
  \theta}{\delta q^\alpha(x)} \frac{\delta }{\delta q^\beta(x)} \;.
\eeq
This can of course only be consistently done, if
\beq{38}
 [\xi_x,\xi_y ] = 0.
\eeq
As in \cite{18} it turns out to be convenient to work with the ``smeared out''
quantities
\beq{39}
 \xi^N_x = \int dx \; N(x) G^{\alpha \beta}(x) \frac{\delta \theta}{\delta
q^\alpha(x)} \frac{\delta}{\delta q^\beta(x) } 
\eeq
with some arbitrary `` test function'' $N(x)$.
We thus get for the commutator
\begin{eqnarray}\label{40}
[\xi^N_x, \xi^M_y ]\!\! & = &
\!\int_x \!\int_y N(x)M(y) G^{\alpha \beta}(x) \frac{\delta \theta}{ \delta
q^{\alpha}(x) } \frac{\delta}{\delta q^{\beta}(x)} G^{\kappa \lambda}(y)
\frac{\delta \theta }{q^\kappa(y)} \frac{\delta}{\delta q^\lambda(y)}
\nonumber \\
& & \!\!\!- \int_x \!\int_y M(x)N(y) G^{\kappa \lambda}(x)
\frac{\delta \theta }{q^\kappa(x)} \frac{\delta}{\delta q^\lambda(x)}
 G^{\alpha \beta}(y) \frac{\delta \theta}{ \delta
q^{\alpha}(y) } \frac{\delta}{\delta q^{\beta}(y)} \nonumber\\
& = & \int_x \!\int_y \left( N(x) M(y) - N(y) M(x) \right) G^{\alpha \beta}(x)
\frac{\delta \theta}{\delta q^\beta(x)} \frac{\delta^2 \theta}{\delta
q^\lambda(y) \delta q^\alpha(x)}    \nonumber      \\
& & \hspace{0.5cm} \times G^{\kappa \lambda}(y) 
    \frac{\delta}{\delta q^\kappa(y) }
\; . 
\end{eqnarray}
This commutator can only vanish if the second functional derivative of
$\theta$ in $(\ref{40})$ is proportional to $\delta(x-y)$.
\medskip

To calculate this quantity we consider the functional version of (\ref{12}),
which reads
\beq{41}
E_x:=\frac{1}{2} G^{\alpha \beta}(x) \frac{\delta \theta}{\delta q^\alpha(x)}
\frac{\delta \theta}{\delta q^\beta(x)} + \frac{1}{\varrho} H \varrho = 0\; .
\eeq
Differentiating $E_x$ with respect to $q^\lambda(y)$ yields
\begin{eqnarray}\label{42}
 0 & = & \frac{\delta E_x}{\delta q^\lambda(y)} + \int dz \;
      \frac{\delta E_x}{\delta (\frac{\delta \theta}{\delta q^\alpha}(z))}
      \frac{ \delta^2 \theta}{ \delta q^\lambda(y) \delta q^\alpha(z) } 
\nonumber \\
   & = & \frac{\delta E_x}{\delta q^\lambda(y)}
      +G^{\alpha \beta} \frac{\delta \theta }{\delta q^\beta(x)}
        \frac{\delta^2 \theta}{\delta q^\alpha(x) \delta q^\lambda(y)}
\; .
\end{eqnarray}
For the first term on the right-hand side we have
\beq{43}
 \frac{\delta E_x}{\delta q^\lambda(y)}  =  F_\lambda \; \delta(x-y) +
 \frac{\delta }{\delta q^\lambda(y)} \frac{1}{\varrho} H \varrho \; ,
\eeq
where the explicit form of the function in front of $\delta(x-y)$ is not
needed below and has therefore been abbreviated by $F_\lambda$.

The Hamiltonian density $H$ is given explicitly by
\beq{44}
 H = - \frac{1}{2M} \, '' \, G_{abcd} \frac{\delta^2}{\delta h_{ab}
\delta h_{cd} } \; ''  -2M \sqrt{h} ( R - 2\Lambda)  + {\cal H}_m \; ,
\eeq
where
\beq{45}
{\cal H}_m = \frac{1}{2} \left( - \frac{1}{\sqrt{h}}
             \frac{\delta^2}{\delta \phi^2}
             + \sqrt{h} h^{ab} \phi,_a \phi,_b
             + \sqrt{h} \left(m^2\phi^2+U(\phi \right) \right)
\eeq
is the Hamiltonian density for a scalar field. Then,
\begin{eqnarray}
\frac{\delta}{\delta q^\lambda(y)} \left( \frac{1}{\rho} H \rho \right)
& = & M_\lambda(x) \delta(x-y) + \frac{\delta V(x)}{\delta q^\lambda(y)} 
\nonumber\\
& = & M_\lambda(x) \delta(x-y) + 2 G^{ijab} \delta,_{ij}(x-y)
      \; \delta_\lambda^{\;\{h_{ab}\}} \nonumber \\
&   & + \sqrt{h} h^{ab} \phi,_a \delta,_b(x-y) \;
       \delta_\lambda^{\; \{\phi \}}  \; , \label{46}
\end{eqnarray}
where here $V$ is given by
\beq{47}
 V = -2 \sqrt{h} ( R-2 \Lambda) + \sqrt{h} h^{ab} \phi,_a \phi,_b
 + \sqrt{h}\left(m^2 \phi^2 + U(\phi)\right) \;,
\eeq
and $\delta_\lambda^{\; \{h_{ab} \}} =1 $ ($\delta_\lambda^{\; \phi}=1$)
if $q^\lambda\in \{h_{ab}\}$ ($q^\lambda=\phi$) and otherwise zero.
The second functional derivatives of the $\delta$-function in (\ref{46})
arise from the Ricci scalar $R$ in (\ref{44}). With the result (\ref{46})
we know the second derivatives of $\theta$ in (\ref{42}), which in turn
give the following expression for the commutator (\ref{40})
\begin{eqnarray*}
[ \xi^N_x, \xi^M_y ]
& = & \int_x  \int_y \; \left( N(x) M(y) - N(y) M(x)\right ) \\
&   & \times  G^{\alpha \beta}(x)\frac{\delta \theta}{\delta q^\beta(x)}
      \frac{\delta^2 \theta}{\delta q^\lambda(y) \delta q^\alpha(x) }
       G^{\kappa \lambda}(y) \frac{\delta}{\delta q^\kappa(y) }  \\
& = & \int_x  \int_y \; \left(N(y)M(x)- N(x)M(y) \right)\\
&  & \times \left[ 2 G^{ijab}(y)
G_{abcd}(y) \delta,_{ij}(x-y) \frac{\delta }{\delta h_{cd}(y)} 
 + h^{ab} \phi,_a \delta,_b(x-y)
\frac{\delta }{\delta \phi(y)} \right] \;.
\end{eqnarray*}
After some partial integrations this yields
\begin{eqnarray}
[ \xi^N_x, \xi^M_y ]
& = & - 2 \!\int_x   ( N \partial_a M - M \partial_a N) \left(
         \frac{\delta}{\delta h_{ab}} \right)_{|b}
         - \int_x  (N \partial_a M - M \partial_a N) h^{ab} \phi,_b
         \frac{\delta}{\delta \phi}
         \nonumber \\
& = & \int dx\; ({\cal L}_K h_{ab})  \frac{\delta}{\delta h_{ab}}
      -  \int dx \; ({\cal L}_K \phi) \frac{\delta}{\delta \phi} \;\;
        \neq 0 \label{48} \;,
\end{eqnarray}
where
\[ K^a := ( N M,_b - M N,_b ) h^{ab} \; .\]
The expressions on the right-hand side of (\ref{48}) are just the
diffeomorphism constraints of general relativity in their quantised form.
Therefore, there is in analogy to the semiclassical case \cite{18} {\it no}
time function $\tau(x)$ available on the space of three-metrics
(since (\ref{48}) does not vanish), but such a function is available on
the space of all three-{\it geometries}, i.e., after the diffeomorphism
constraints are divided out.
This of course makes sense, since it is assumed that in the inner
product which is used for the Ehrenfest equations all unphysical
variables are eliminated.

Note that for our result only the special structure of the time derivative
(\ref{37}) and the functional version of equation (\ref{12}) as a condition
for the phase of the wave function are important. For this reason every  
description by ``time vector fields'' with a representation (\ref{48}) 
has this property. {}From this point of 
view the analogous result for the WKB time (\ref{30}) in \cite{18} comes 
out naturally.
We can also conclude immediately that the same result must hold for every 
order of the semiclassical approximation scheme described in \cite{7}, 
if the higher orders
are understood to describe the influence of back reaction on $t_{W\!K\!B}$ 
by corrections to the phase.
\medskip

After this test of consistence, we finally show a characteristic property
of the transformation from the  old variables (the three-metric and
matter fields, collectively denoted by $\{q^\alpha \}$ ) to the Ehrenfest
variables.

For this purpose we insert into (\ref{41}) the functional form of
$G'^{00} \dot{\theta} =1$ (cf.(10)), which reads
$ G^{\alpha \beta}\frac{\delta
\theta}{\delta q^\alpha} \frac{\delta \theta}{\delta q^\beta}= G'_{00}$
and find
\beq{49}
  G'_{00}= -\frac{2}{\varrho} H \varrho \;.
\eeq
Assuming that the metric depends on the new coordinates ultralocally (i.e., 
that it contains no spatial derivatives of the coordinate),
partial differentiation of (\ref{49}) leads to
\[ \frac{\delta G'_{00}(x)}{\delta q^\alpha(y)} = \int dz \;
      \frac{\delta \eta^\sigma(z)}{\delta q^\alpha(y)}
      \frac{\delta G'_{00}(x)}{\delta \eta^\sigma(z)}
    \equiv \frac{\delta \eta^\sigma(x)}{\delta q^{\alpha}(y)} K_\sigma(x) \;.
\]
Comparing this with (\ref{46}) yields the following expressions for the
coordinate transformation
\begin{eqnarray*}
\frac{\delta \eta^\sigma(x)}{\delta h_{ab}(y)} K_\sigma(x) & = &-4 G^{ijab}
\delta,_{ij}(x-y) + ... \delta(x-y)   \; , \\
\frac{\delta \eta^\sigma(x)}{\delta \phi(y)} K_\sigma(x) & = &
-2 \sqrt{h} h^{ab} \phi,_a \delta,_b(x-y) + ... \delta(x-y) \;.
\end{eqnarray*}
We recognise from these expressions that the transformation from the old
coordinates to Ehrenfest coordinates {\it cannot be ultralocal}.


\section{Summary and Conclusion }
\bigskip

In this paper we have studied various consequences which arise in using
the Ehrenfest equations as a way to solve the problem of time in quantum
gravity [10-13]. Is the Ehrenfest time a viable candidate for a concept of
time in quantum gravity?

In the following we investigate this concept in view of the ``problems of
time'' which are listed in \cite{2}.
\medskip

{\it Existence problem:} This may in fact be the major problem. As we have
shown in Section 2, only very few
solutions of the Wheeler-DeWitt equation allow the validity of the Ehrenfest
condition (\ref{10}) (with $\kappa=1$) and (\ref{17}). Although it was
of course clear that the proposal does not work for real solutions (such
as, e.g., the Hartle-Hawking wave function), this drastic restriction does
not seem to have been considered in [10-13]. We emphasise that the existence
problem already occurs locally  in configuration space, independent
of possible global obstructions which one would expect to arise anyway.

{\it Hilbert space problem}: This concerns the question of the correct
inner product in quantum gravity. {\it If} the Ehrenfest equations hold,
this problem is solved by the choice made in (\ref{2}).
Moreover, through equation (\ref{1}) it is possible to get sensible
answers for  expectation values of observables from
the ``wave function of the universe''.

{\it Uniqueness problem}: In case of existence, one can uniquely construct the 
corresponding phase time from a given solution to the Wheeler-DeWitt equation 
after an initial hypersurface $t_0 = const.$ has been specified.

The {\it spacetime  problem} as well as the {\it sandwich problem} do not
play any role in this approach, since they only arise if embeddings
of hypersurfaces into spacetime are considered. But there is a new problem,
the {\it semiclassical problem:} Only very few solutions which allow an
Ehrenfest time do possess a sensible semiclassical limit, as was shown
in Section 3.
\medskip

Thus, in summary, the alternatives are either to reject this proposal as
a solution to the problem of time in quantum gravity, or to interpret it as
an additional {\it boundary condition} to extract a sensible solution
from the Wheeler-DeWitt equation. 
It is important in this respect to note that the Ehrenfest proposal does not
respect the superposition principle, i.e., the sum of two ``Ehrenfest
solutions'' is not  an Ehrenfest solution again.
Whether such an
``Ehrenfest boundary condition'' turns out to be successful is an issue which
has not yet been explored.


\newpage
\begin{appendix}
\section{Appendix: Calculation of the second Ehrenfest condition}
\bigskip

In this appendix we shall present the necessary steps to derive Eq.
(\ref{17}).
For this purpose we first calculate $\tilde{V}$, cf. (\ref{7}),
\begin{eqnarray}\label {A.1}
\widetilde{V} & = & \frac{1}{\Psi^*\Psi}\left[ \frac{1}{2} Re(\Psi^* D^2 \Psi)
- Im (\Psi^*\partial_t \Psi) \right] \nonumber \\
& = & \; \frac{1}{\varrho^2} \;\left[-\frac{1}{2}Re(\Psi^* D_0^2 \Psi)
+V\varrho^2 + \xi {\cal R} - \varrho^2 \dot{\theta} \right] \; .
\end{eqnarray}
Inserting $D_0^2$ and calculating the real part leads to
  \[
V = \widetilde{V} -\xi {\cal R}+ \dot{\theta} - \frac{1}{2} G'^{00} 
\dot{\theta}^2 +
\frac{1}{2} \frac{1}{\varrho \sqrt{G'} } \partial_t(\sqrt{G'} G'^{00}
\dot{\varrho} )\; .
\]
The first step in our derivation is given by the following
\medskip

\noindent
{\it Proposition:}
\[ \frac{d}{dt}<p_i> = i<\! [\widetilde{H},p_i] \! > \;,\]
where $p_i$ is given in (\ref{15}), and $\widetilde{H} \equiv -\frac{1}{2} 
D^2 +\widetilde{V}$.
\medskip

\noindent
{\it Proof:}
\[<i[\widetilde{H}, p_i]> \hspace{12cm} \]
\vspace*{-0.75cm}
\begin{eqnarray*}
\hspace{0.2cm}
& = & i\! \int\!\! D \eta \sqrt{G'} \left[ (\widetilde{H} \Psi)^* p_i \Psi
    -\Psi^* p_i \widetilde{H} \Psi \right] \\
& = &  \int\!\!D \eta \sqrt{G'} \left[ \frac{1}{2} 
    (\partial_t\!\ln\!\sqrt{G'}) \Psi^*
    p_i \Psi + \partial_t  \Psi^*  p_i \Psi + \Psi^*
    p_i \partial_t  \Psi + \frac{1}{2} \Psi^* p_i  
   (\partial_t\!\ln\!\sqrt{G'}) \Psi \right] \\
& = &  \int \!\!D \eta \sqrt{G'} \left[   (\partial_t \ln\!\sqrt{G'})  \Psi^*
    p_i \Psi + \partial_t  \Psi^* \, p_i \Psi + \Psi^*
    p_i \partial_t  \Psi -  \Psi^* \frac{i}{2}(\partial_i \partial_t
    \ln\!\sqrt{G'})\Psi  \right] \\
& = &  \int\! \!D \eta \sqrt{G'} \left[ \; (\partial_t \ln\!\sqrt{G'}) 
    \; \Psi^*
    p_i \Psi + \partial_t \Psi^* \, p_i \Psi + \Psi^*
    p_i \partial_t  \Psi + \Psi^*
    \frac{\partial p_i}{\partial t} \Psi \right] \;.\\
\end{eqnarray*}
The last equality follows from
\beq{A.2}
\frac{\partial p_i}{\partial t} 
 = -\frac{i}{2} \partial_t (\partial_i +\frac{\partial_i 
      \sqrt{G'}}{\sqrt{G'}}  ) 
 =  -\frac{i}{2}\partial_t(2 \partial_i + (\partial_i \ln\sqrt{G'}))
 =  -\frac{i}{2} ( \partial_i \partial_t \ln \sqrt{G'} ) \;.
\eeq
Thus, our proposition has been proven.
\medskip

Since the expectation value of the explicit time dependence of $p_i$ must
be real, it is clear from (\ref{A.2}) that
\beq{A.3}
 < \frac{\partial p_i}{\partial t} > = 0 = <\partial_i \partial_t \ln 
\sqrt{G'} >
\eeq
must hold. The second Ehrenfest equation is thus fulfilled if
\[  < \!i [\widetilde{H}, p_i] \! >= <\!i [H,p_i] \! > \;.\]
To show this we first insert the expressions for $H$, $\widetilde{H}$ and
$\widetilde{V}$ into this equation:
\beq{A.4}
  <\! i [-\frac{1}{2} D_0^2 + \dot{\theta} - \frac{1}{2} G'^{00} \dot{\theta}^2
  + \frac{1}{2} \frac{1}{\varrho \sqrt{G'}} \partial_t( \sqrt{G'} G'^{00}
  \dot{\varrho} ) , p_i ] \!>  = 0 \; . 
\eeq
We first determine
\begin{eqnarray*}
2i\left[ D_0^2, p_i \right] \Psi  & = & \left[\frac{1}{\sqrt{G'}} \partial_t
\sqrt{G'} G'^{00} \partial_t, \partial_i + \frac{1}{\sqrt{G'}} \partial_i
\sqrt{G'} \right] \Psi \\
& = & \frac{2}{\sqrt{G'}} \partial_t \sqrt{G'} G'^{00} \partial_t \partial_i
      \Psi
   - \frac{2}{\sqrt{G'}} \partial_i \partial_t \sqrt{G'} G'^{00} \partial_t
      \Psi \\
&   & + \frac{1}{\sqrt{G'}} (\partial_i \ln \sqrt{G'} ) \partial_t \sqrt{G'}
      G'^{00} \partial_t \Psi \\
&   & + \frac{1}{\sqrt{G'}} \partial_t \sqrt{G'} G'^{00} (\partial_t
       \partial_i \ln \sqrt{G'} ) \Psi \\
&   & + \frac{1}{\sqrt{G'}} \partial_t \sqrt{G'} G'^{00} (\partial_i \ln
      \sqrt{G'}) \partial_t \Psi \\
& = & 2 G'^{00} \partial_t^2 \partial_i \Psi + \frac{2}{\sqrt{G'}}
      (\partial_t \sqrt{G'} G'^{00} ) \partial_t \partial_i \Psi \\
&   & - \frac{2}{\sqrt{G'}} \partial_i \sqrt{G'} G'^{00} \partial_t^2 \Psi
      - \frac{2}{\sqrt{G'}} \partial_i (\partial_t \sqrt{G'} G'^{00})
       \partial_t \Psi \\
&   & + (\partial_i \ln \sqrt{G'}) G'^{00} \partial_t^2 \Psi
      + \frac{1}{\sqrt{G'}} (\partial_i \ln \sqrt{G'}) ( \partial_t \sqrt{G'}
      G'^{00}) \partial_t \Psi \\
&   & + G'^{00} ( \partial_t \partial_i \ln \sqrt{G'}) \partial_t \Psi
      + \frac{1}{\sqrt{G'}}(\partial_t \sqrt{G'} G'^{00} 
      ( \partial_t \partial_i
      \ln \sqrt{G'})) \Psi \\
&   & + G'^{00} ( \partial_i \ln \sqrt{G'} ) \partial_t^2 \Psi
      + \frac{1}{\sqrt{G'}} ( \partial_t \sqrt{G'} G'^{00} (\partial_i \ln
      \sqrt{G'})) \partial_t \Psi \\
& = & - 2(\partial_i G'^{00}) \partial_t^2 \Psi
 -\frac{2}{\sqrt{G'}}(\partial_t \sqrt{G'} (\partial_i G'^{00}))
      \partial_t \Psi \\
&   & +\frac{1}{\sqrt{G'}} ( \partial_t \sqrt{G'} G'^{00} 
      (\partial_t \partial_i
      \ln \sqrt{G'})) \Psi  \\
& = & - 2(\partial_i G'^{00}) [\frac{\ddot{\varrho}}{\varrho} + 2i
      \frac{\dot{\varrho}}{\varrho} \dot{\theta} + i \ddot{\theta} -
      \dot{\theta}^2] \Psi \\
&   & -\frac{2}{\sqrt{G'}}(\partial_t \sqrt{G'} (\partial_i G'^{00}))
      [\frac{\dot{\varrho}}{\varrho} + i\dot{\theta} ] \Psi \\
&   & +\frac{1}{\sqrt{G'}} ( \partial_t \sqrt{G'} G'^{00} 
      (\partial_t \partial_i \ln{\sqrt{G'}})) \Psi \;.
\end{eqnarray*}
Then we calculate
\[
2i[V+ \xi {\cal R}-\widetilde{V}, p_i] \Psi =  -2(\partial_i \dot{\theta}) \Psi
          + (\partial_i G'^{00} \dot{\theta}^2) \Psi
          - (\partial_i \frac{ ( \partial_t
           \sqrt{G'} G'^{00} \dot{\varrho}) }{\varrho \sqrt{G'}})\Psi \; .
\]
Inserting these results into (\ref{A.4}) and performing a decomposition
into real and imaginary part yields
\begin{eqnarray*}
0 & = & \int D\eta \; \sqrt{G'}\varrho^2 \left[- \frac{( \partial_t \sqrt{G'}
      (\partial_i G'^{00})\dot{\varrho}) }{\sqrt{G'}\varrho} 
  +\frac{(\partial_t \sqrt{G'}G'^{00} (\partial_t \partial_i 
   \ln{\sqrt{G'}}))}{2\sqrt{G'}} \right.\\  
  &  & \hspace{2.5cm} \left.+ (\partial_i \frac{1}{\varrho \sqrt{G'}} 
    (\partial_t \sqrt{G'} G'^{00}
  \dot{\varrho})) +2(\partial_i \dot{\theta}) - G'^{00} (\partial_i
  \dot{\theta}^2) \right]\;,\nonumber
\end{eqnarray*}

\[ 0=\int D\eta \; \partial_t\left(\sqrt{G'}\varrho^2 
     (\partial_i G'^{00})  \dot{\theta}\right) \;.\]
We now make use in these equations of the relations $\theta = \theta(t)$,
$G'^{00} \dot{\theta} = \kappa(\eta)$ and $\partial_t(\sqrt{G'} \varrho^2)=0$,
which follow from the first Ehrenfest equation. One immediately recognises that
the imaginary part vanishes. The vanishing of the real part can be explicitly
checked as follows:
\[
- \frac{( \partial_t \sqrt{G'} (\partial_i G'^{00})\dot{\varrho}) }{\sqrt{G'}
         \varrho} 
  +\frac{(\partial_t \sqrt{G'}G'^{00} (\partial_t \partial_i 
   \ln{\sqrt{G'}}))  }{2\sqrt{G'}}
     + \partial_i \frac{(\partial_t \sqrt{G'} G'^{00}
  \dot{\varrho}) }{\varrho \sqrt{G'}} \hspace*{2.5cm}
\]
\vspace*{-0.75cm}
\begin{eqnarray*}
 &=& -\varrho(\partial_t(\partial_i G'^{00}) \frac{\dot{\varrho}}{\varrho^2})
    + \frac{1}{2} \varrho^2 (\partial_t \frac{G'^{00}}{\varrho^2}
      \partial_t \partial_i\ln \sqrt{G'}) 
    + \partial_i(\varrho \partial_t G'^{00} \frac{\dot{\varrho}}{\varrho^2}) \\
 &=& -\partial_t(\varrho(\partial_i G'^{00}) \frac{\dot{\varrho}}{\varrho^2})
    + (\partial_i G'^{00}) \frac{\dot{\varrho}^2}{\varrho^2}
    - \partial_t(G'^{00} \partial_i(\frac{\dot{\varrho}}{\varrho})) \\
& & +2 \frac{\dot{\varrho}}{\varrho} G'^{00} \partial_i
       (\frac{\dot{\varrho}}{\varrho        })
    + \partial_i \partial_t (G'^{00}\frac{\dot{\varrho}}{\varrho})
    - \partial_i(G'^{00} \frac{\dot{\varrho}^2}{\varrho^2}) \\
&=&  \partial_t(G'^{00} \partial_i(\frac{\dot{\varrho}}{\varrho})) 
    -  G'^{00} \partial_i(\frac{\dot{\varrho}^2}{\varrho^2 })
    - \partial_t(G'^{00} \partial_i(\frac{\dot{\varrho}}{\varrho})) 
   +2 \frac{\dot{\varrho}}{\varrho} G'^{00} \partial_i
      (\frac{\dot{\varrho}}{\varrho        }) \\
&=& 0 \;.
\end{eqnarray*}
Thus, (\ref{A.4}) holds, and the only condition from the second Ehrenfest 
equation is (\ref{A.3}). We note that the factor ordering term $\xi {\cal R}$
did not play any role in the derivation of (\ref{A.3}). Furthermore, the 
exact form of $\kappa(\eta)$ (which we have fixed to be $\kappa\! \equiv \!1$
from semiclassical considerations) does not enter these calculations.

That the imaginary part of (\ref{A.4}) must vanish follows of course
immediately from the fact that $\tilde{H}$ and $p_i$ are both hermitean
and that thus $<\!i[\tilde{H}-H,p_i]\!>$  must be real.
\end{appendix}


\end{document}